\documentclass[prb,twocolumn,showkeys,preprintnumbers,amsmath,amssymb]{revtex4}

\usepackage{graphicx}
\usepackage{dcolumn}
\usepackage{bm}

\begin{document}


\title{Colloquium: Comparison of Astrophysical and Terrestrial Frequency Standards}


\author{John G. Hartnett and Andre N. Luiten}
\affiliation{School of Physics,  University of Western Australia,
    Crawley, WA 6009, Australia}
\email{john@physics.uwa.edu.au}


\begin{abstract}
We have re-analyzed the stability of pulse arrival times from pulsars and white dwarfs using several analysis tools for measuring the noise characteristics of sampled time and frequency data.  We show that the best terrestrial artificial clocks substantially exceed the performance of astronomical sources as time-keepers in terms of  accuracy (as defined by cesium primary frequency standards) and stability. This superiority in stability can be directly demonstrated over time periods up to two years, where there is high quality data for both. Beyond 2 years there is a deficiency of data for clock/clock comparisons  and both terrestrial and astronomical clocks show equal performance being equally limited by the quality of the reference timescales used to make the comparisons. Nonetheless, we show that detailed accuracy evaluations of modern terrestrial clocks imply that these new clocks are likely to  have a stability better than any astronomical source up to comparison times of at least hundreds of years.  This article is intended to provide a correct appreciation of the  relative merits of natural and artificial clocks. The use of natural clocks as tests of physics under the most extreme conditions is entirely appropriate; however, the contention that these natural clocks, particularly white dwarfs, can compete as timekeepers against devices constructed by mankind is shown to be doubtful.
\end{abstract}


\maketitle

\section{Introduction}
It has been asserted numerous times in the academic literature that pulsed astrophysical sources could be,~\citep{Allan1987,Rawley1987,sazhin1989,allan1989,Hama1989, petit1996, matsakis, Taylor1991, Manchester2004,Roy2005,Jacoby2005,Hotan2006,Zhong2007,Verbiest2009} or are~\citep{Mukadam2003,Kepler2005,Astro303}  the best clocks in the Universe.  This belief is even more widely perpetuated in the popular science literature. 
 In this article we contend that this assertion is  not correct and, in fact,  today, by all sensible measures of clock performance, artificial clocks have substantially better performance than natural astrophysical clocks, at least 
out to timescales of a few years.  We believe that though there may have been some basis for these claims in the past, this is no longer the case, and in some cases, the claims have been based on an erroneous analysis leading to an incorrect comparison of the relative stability of astronomical pulse arrival times and fluctuations  in the frequency of artificial clocks.  

In this paper we present formulas for making comparisons between artificial clocks, white dwarfs and pulsars and also present new analysis of several millisecond pulsars together with an analysis of a set of astrophysical sources which have been represented as   having the potential to function as high performance clocks. However, it should be noted that in all cases where we refer to pulsar (or white dwarf) stability we are actually referring to the comparison of the astrophysical clock signal against a terrestrial reference clock and/or timescale developed from an ensemble of many atomic clocks, for example, TAI and TT(BIPM). \cite{TT(BIPM)}

We make use of two types of statistics for estimating the timekeeping stability: first, the Square Root Allan Variance (SRAV), represented by $\sigma_y(\tau)$ (also called the Allan Frequency Deviation in more modern literature), which is frequently used in the time and frequency community.  The second  is the $\sigma_z(\tau)$ statistic,~\citep{matsakis} which finds popularity in the pulsar timing community. 

The $\sigma_z$ statistic is used as the quantifier of pulsar time of arrival (TOA) signals that, among other things, is insensitive to the deterministic but \textit{a priori} unknown average period drift $\dot{P}$ (or first derivative of pulse frequency) while allowing one to study the long-term random noise characteristics of the data. This has been found to be suitable when studying millisecond pulsars over the longest time periods---a ``pulsar variance.'' 
The outcome of that analysis has resulted in some pulsars being referred to as ``nature's most stable clocks''.~\citep{Taylor1991} 

We also show that  $\sigma_y(\tau)$ with period drift removed and $\sigma_z(\tau)$ yield comparable values over $\tau \approx 10$ year time periods for the best pulsars (measured against atomic clocks). For all periods $\tau < 3$ years, $\sigma_y(\tau)$, with period drift removed, gives a lower noise statistic than does $\sigma_z(\tau)$. Except over the longest measured time periods the pulsar data, shown in this paper, are thus more favorably represented by $\sigma_y(\tau)$. 

For the pulsars we have thus removed a measured yet \textit{a priori} unpredictable average period drift $\dot{P}$ from the TOA data, which results in the most optimistic statistic for these pulsars.~\citep{Verbiest2008, Verbiest2009}  If a significant level of period drift $\dot{P}$  is present, and not removed before analyzing the random contributions, then it will contribute to the total instability with a $(|\dot{P}|/P)(\tau/\sqrt{2})$ dependence. In the case of  modern terrestrial clock/clock comparison data it is unnecessary to remove any such effect.  And for the very longest averaging times we compare $\sigma_y(\tau)$ and $\sigma_z(\tau)$ for the millisecond pulsar PSR J0437-4715.

The square root Allan variance  can be calculated from either the fractional frequency fluctuations of the oscillator or from the period fluctuations in the form of timing residuals, both measured with respect to a lower noise reference. In the 1980s a comparison was made along similar lines~\citep{Cordes1985, Ilin1989} in which it was  found that one pulsar, a millisecond pulsar (PSR B1937+21), rivaled the frequency stability of the hydrogen-maser (H-maser).  Those workers concluded, however, that most pulsars were much noisier than a H-maser.  

There has also been an examination of the potential use of ensemble pulsar timing data to correct the very long term fluctuations of terrestrial timescales. Based on the quality of the timescales of that era it was suggested that this could work with the discovery of new high-stability pulsars and  longer-term observations.~\citep{matsakis96,petit1996} However, since that time there has been a substantial improvement (by up to two orders of magnitude) in the accuracy and stability of artificial terrestrial clocks. There have also been new claims that  pulsating white dwarfs have a superior stability to even the best millisecond pulsars.~\citep{Mukadam2003,Kepler2005}  

Thus, in this paper we hope to present two important additions to the literature: that the improvement in modern terrestrial clock performance means that it is doubtful that astrophysical sources would be able to correct terrestrial timescales on periods less than 10 years, and, in addition, that a re-analysis of these stability of these pulsating white dwarfs actually shows that they are substantially inferior to the millisecond pulsars. We believe it is thus a sensible and opportune time to revisit the claims being made for high quality natural clocks together with a comparison to modern terrestrial clocks.

\subsection{Definition of the SI second}
To commence, one should clearly differentiate between the concept of accuracy and stability. On occasion, the word \textit{accuracy} has been loosely applied to the stability of a pulse train delivered by natural clocks such  as pulsars.~\citep{Time1968,Thomsen1985, Bronnikov1989, Astro303, Lyne2006, Zhong2007, Karastergiou2007} 
 Any attempt to produce an \textit{accurate} time standard requires that the frequency of the oscillator can be related to the definition of the SI second, which at the present time is    defined in terms of a hyperfine splitting in the ground state of the cesium atom.~\citep{SIsecond}   It is possible that in the future this SI definition may be replaced with another, although any   new definition is sure to be closely connected to the values of the universal and presumed unchanging fundamental constants.~\citep{TaylorB}

Contrasting with this universal and unchanging definition, the frequency  of a natural astrophysical source is determined by some stochastically distributed initial period together with some additional processes that have occurred in the evolution of the star to its current observed rotational status.  It is thus extremely unlikely that these natural pulsating sources could  ever be the basis of an \textit{accurate} time system.~\citep{Riehle, petit1996, matsakis1996,matsakis96} Furthermore, as we will see below, one can perform a careful accuracy evaluation on terrestrial clocks in order to ascertain all potential sources of systematic uncertainty. This is something that is not possible with the astronomical sources and, as we will show, allows us to extrapolate the frequency stability of the clocks to time periods over which it hasn't yet been possible to measure. This allows us to place a limit on the worst possible performance in that integration time regime.

\subsection{Pulsar timescale}
In spite of this intrinsic inaccuracy, one can still conceive of a  pulsar timescale, derived from an agreed-upon ensemble of pulsars, which might prove to be very useful if it were to provide a universally accessible and highly \textit{stable} timescale over the longest times ($> 10$ years).~\citep{matsakis96,petit1996} In light of this possibility  it is definitely worthwhile  examining the potential stability of these astrophysical sources. These pulsar timing arrays could potentially verify the extreme long term performance of terrestrial timescales.~\citep{matsakis1996}

Modern artificial clocks have undergone detailed frequency accuracy evaluations~\cite{footnote1}  at the level of a few parts in $10^{17}$.  The implication of these evaluations is that the frequency stability will, over some sufficiently long measurement time, exhibit a stability which is at or below this same value. Once the frequency stability has passed under this accuracy evaluation limit it will not exceed this value for arbitrarily long measurement times. On the other hand, the best pulsars, measured by comparison with local reference clocks, have relatively poor signal to noise leading to a stability of $\sigma_z(\tau)\approx 10^{-15}$ for $\tau \approx 10$ years.~\citep{Verbiest2008, Verbiest2009}  Therefore only by averaging the pulsar pulse timing sequence for times possibly greater than hundreds of years  is it possible for a pulsar pulse sequence to present a superior stability to that of the best earth-based clock.  We reiterate that what is currently measured is actually the stability of the pulsar compared to some terrestrial local atomic reference timescale rather than direct pulsar-pulsar comparisons.

This article and analysis is not intended to undermine the potential for pulsars to be excellent tests of cosmology, general relativity and astrophysics in extreme conditions where they are  the most sensitive tests of these type of physics yet performed.~\citep{Stairs2004, Lattimer2004, Lyne2004, Manchester2004, Kramer2006, Breton2008}  

\section{Frequency stability and accuracy of clocks}
The conventional measure which metrologists use to characterize the stability of any oscillator is the SRAV.~\citep{allan1966, barnes1971} The measure operates on a time series of frequency measurements taken over some integration time, $\tau$, and is usually quoted over a range of different integration times to capture the behavior of the oscillator over these various time periods. The SRAV is a close relative of the more conventional standard deviation but is stable in the presence of various divergent noise types that afflict real oscillators.~\citep{CCIR1986, Sullivan}  In practice the  SRAV, $\sigma_{y}(\tau)$, is calculated from a time series of $N$ frequency measurements, $f_{n},f_{n+1},f_{n+2}...$, where, 
\begin{equation}
\sigma_{y}^{2}(\tau)= \frac{1}{2 N \nu_{0}^{2} }  \sum_{n=0}^{N-1} \left( f_{n+1}-f_{n} \right)^{2},
\end{equation}
where $\nu_{0}$ is the nominal average frequency of the oscillator. The SRAV is defined assuming that the integration time equals the time between samples; for other circumstances there are published corrections.~\citep{Dawkins} Following~\citep{barnes1971} one an also determine the SRAV from,
\begin{eqnarray}
& \sigma_{y}^{2}(\tau)= \frac{1}{16 \pi^{2} \tau^{2} \nu_{0}^{2} (N-1) } \times  ... \nonumber \\
& \times \sum_{n=0}^{N-2} \left[  \phi(t_{n+2})-2 \phi(t_{n+1}) +  \phi(t_{n})  \right]^{2},
\end{eqnarray}
where $\phi(t)$ is the phase of the oscillator at some epoch $t$, and where in this case  $\tau$ is defined as the time between samples. In the case of a pulsed source (such as a pulsar) we can rewrite Eq. (2) in terms of  timing errors as,
\begin{eqnarray}
& \sigma_{y}^{2}(\tau)= \frac{1}{4   \tau^{2}  (N-1) } \times ... \nonumber \\
& \times \sum_{n=0}^{N-2} \left[X(t_{n+2})-2 X(t_{n+1}) + X(t_{n})  \right]^{2},
\label{eq3}
\end{eqnarray}
where $X(t)$ is defined as the time difference (measured in seconds) between the actual epoch of some well-defined  event and the predicted epoch of that same event according to some ideal clock.  

The subtlety inherent in all the measurement procedures described above is that one must always compare the frequency of some clock-under-test against another reference clock of equal or better performance if one wishes to be confident about the results of the measurement: this is just as true for natural as it is for artificial clocks.  

\subsection{Astrophysical and terrestrial frequency standards compared}
\begin{figure}[!t]
\centering
\includegraphics[width=3.5in]{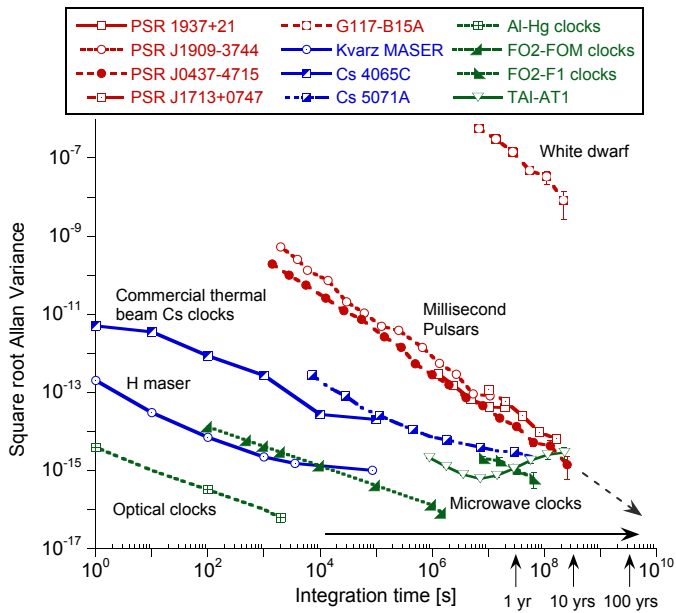}
\caption{\label{fig1} Derived frequency stability for a range of different frequency sources including astrophysical sources (shown in red in online version), commercial oscillators (shown in blue) and the best laboratory clocks (shown in green).  See text for a complete description. For clarity, errors have been included only on the most relevant data to the discussion. The solid line arrow indicates the upper limit of the long-term atomic clock stability due to the known accuracy  of one of the optical ion clocks at $2.3 \times 10^{-17}$. The dashed line arrow shows the long-term trend for the best millisecond pulsar. }
\end{figure}

The definition for the Allan Variance in Eq. (3) above is very close to the usual method for measuring the temporal stability of the pulse train  emitted by a white dwarf or a pulsar detected on earth. For example, in Kepler et al. (2005), the instant of the brightest light emission of a  white dwarf (G117-B15A) with a 215.197s period is measured irregularly over $\sim 30$\,years.  We have re-analyzed their data and presented it in terms of the SRAV on Fig.~\ref{fig1}. In addition, on this plot we show $\sigma_y$ calculated  from TOA data for  three millisecond pulsars (PSR J1909-3744, PSR J0437-4715 and PSR J1713+0747),~\citep{Hotan2006, Verbiest2008, Verbiest2009} as well as one (PSR B1937+21) which was first proposed many years ago as a potential timekeeper.~\citep{allan1989}  The timing residuals used here for PSR J0437-4715 are shown in Verbiest et al. (2008)~\citep{Verbiest2008} and those for PSR J1909-3744 and PSR J1713+0747 are shown in Verbiest et al. (2009).~\citep{Verbiest2009} Both of these papers describe in some detail the analysis performed to obtain them and we have followed exactly the same procedure.  The SRAV shown here has been calculated from those timing residuals after linear period drift was removed.

Figure~\ref{fig1} presents the SRAV of a selection of the best artificial clocks and frequency standards derived from measurement procedures as outlined above.  The line labeled \emph{Al-Hg clocks} represents a comparison of the  Al$^+$ and  Hg$^+$ ion clocks at NIST, USA, which are both based on optical interrogations of laser cooled and  trapped ions. The  result reported here is effectively a measurement of the combined stability of both clocks as they have similar stability. 

The line labeled FO2-FOM clocks shows the best performing microwave frequency standards and is a comparison of a Rubidium fountain clock and the mobile Cesium fountain from the Observatory of Paris (SYRTE).~\citep{guena2008} The line labeled FO2-F1 clocks is derived from a 6 year comparison between the atomic fountain clocks at NIST (F1) and at SYRTE (FO2).~\citep{Parker1} The data are somewhat sparse so we do not present stability data beyond 2 years integration time.  It is important to be aware that NIST F1 and SYRTE FO2 clocks are similar in performance.  The larger instability in the FO2-F1 data comes from the comparison process, due to significant frequency transfer uncertainty, imperfect alignment of runs, and a high rate of dead time.~\citep{Parker2}   

We show the specified frequency stability of a commercial H-maser (labeled as Kvarz MASER),~\citep{kvarz} which we have confirmed in comparison with a UWA cryogenic sapphire oscillator.~\citep{Hartnett2006}  The  performances of commercial  thermal beam cesium clocks are shown by the curves labeled Cs 4065C~\citep{Cs4065C} and Cs 5071A.~\citep{Cs5071A} In the latter a particular cesium 5071A standard was compared with the NIST clock ensemble over 7 years until  the cesium supply was fully depleted. The time residual data are spaced every 2 hours (with very few missing) from which the SRAV was calculated after linear drift was removed. This represents the entire life of a high performance Cs tube. 

Finally, in the line labeled TAI-AT1 we show $\sigma_y$ calculated for the inter-comparison of two terrestrial timescales, from the time difference between International Atomic Time (TAI) and a free running NIST clock ensemble (called AT1), after linear drift was removed.  We note that these timescales are based on hundreds of Cs atomic clocks and hydrogen masers and that the longevity of a timescale is not dependent upon the type, model, or longevity of any individual device. For example, the gradual introduction of high-performance hydrogen masers into AT1 over the last decade has resulted in an improved stability for that timescale. 

The principal feature to note from this plot   is the   much higher  performance of artificial oscillators/clocks in comparison with the astrophysical sources over short measurement times.  In addition, in strong contrast with the claims by Kepler et al.,~\citep{Kepler2005} the white dwarf stability is substantially inferior  when compared with both the millisecond pulsars and the artificial clocks.
 
It can also be noted that the slopes of the various curves on the plot are not all the same.  In the case of the astrophysical sources one sees mostly a  slope of $1/\tau$ which is characteristic of a source that is dominated by white phase noise.~\citep{Rutman1978} This is what would  be expected in the case of a measurement   based on measuring the  epoch of a particular event in the presence of white noise (principally receiver noise because of the limited signal to noise ratio). 

Most of the atomic clocks show a downward slope of $1/\sqrt{\tau}$, which is characteristic of the presence of white frequency noise,~\citep{Rutman1978} that arises when  locking an oscillator to a frequency dispersive feature in the presence of white detection noise.  The H-maser and Cs beam clocks do show some evidence of different slopes at the longest times associated with quadratic frequency drift and random walk of frequency.  

Although, the longest terrestrial measurements trend upwards for the longest integration times to nearly meet the best of the pulsars, one should emphasize that these are known limitations of these older standards.  There is no evidence (over their necessarily shorter measurement times) of frequency drift from the SRAV calculated for any of the other modern atomic clocks.
 
One notes that the FO2-F1 clock comparison has an instability characterized by $\sigma_y \approx 6 \times 10^{-12} \tau^{-1/2}$ over the integration times shown; although, as mentioned above, this is not indicative of the clocks themselves but instead due  to technical issues associated with comparing widely separated clocks on different continents. Similarly pulsars seen from different  hemispheres need local reference clocks and to make pulsar/pulsar comparisons of such the local clock instabilities also afflict those measurements. \cite{footnote2} Nevertheless, at its closest approach this clock comparison is still an order of magnitude below the best pulsars, and three orders of magnitude better, if we extrapolate the SRAV  of the underlying Cs fountain stability (as exemplified by the FO2-FOM clock comparison) so as to eliminate the added uncertainties which resulted from the comparison as explained above. 

\subsection{Accuracy evaluation}
In order to obtain a worst-case estimate of the stability of modern atomic clocks for long-term integration times, which have not been directly measured,  we examine recently performed accuracy evaluations on these devices.  The accuracy evaluation encapsulates  all possible systematic uncertainties of the device and could be interpreted as the potential performance of a timescale if it were to be based on such a device.  The accuracy evaluation can be also regarded as the maximum fractional deviation of the clock output frequency from the ``true'' underlying and unperturbed atomic transition frequency. Thus one sees that, given sufficient integration to average away white noise associated with the measurement process, it is clear that the frequency instability, measured between two clocks with a certain accuracy evaluation, cannot exceed the bound set by the accuracy evaluation itself.

The most recent accuracy evaluations have been performed for the Hg and Al optical ion clocks~\citep{rosenband2008} at the level of 1.9 and $2.3 \times 10^{-17}$ respectively, the Sr optical clock~\citep{Ludlow,Campbell, Akatsuka} at $1-1.4 \times 10^{-16}$,  the Yb optical clock~\citep{Lemke} at $3.4 \times 10^{-16}$, and the Cs and Rb fountain clocks at SYRTE~\citep{Chapelet,guena2008} at the level of around $4.5 \times 10^{-16}$.   It should be noted that the Hg and Al optical ion clock accuracy evaluations are below that of the current realization of the SI second, hence, they have actually reported the ratio of their optical frequencies, so as to avoid the uncertainty ($3.3 \times 10^{-16}$) of the currently realized SI second.~\citep{Jefferts2007}  The same is also true for the Sr lattice clock in which a remote optical comparison was made with an atomic Ca clock.~\citep{Ludlow}

As an example of the power of the systematic uncertainty evaluation, the particular Cs 5071A beam clock used to make the comparison, shown as Cs 5071A in Fig. 1, has an accuracy~\citep{Parker2010} of $2 \times 10^{-15}$  and one sees that its stability approaches that value at the longest times. One can also see that the stability of the Al-Hg ion clocks comparison also approaches its measured accuracy evaluation at the longest integration times, indicated by a solid arrow in Fig. 1.  We note that commercial Cs beam atomic clocks may be more stable over 1 year of averaging than over 5 years, but nonetheless their long-term stability always remains below their specified systematic uncertainty (which sets the upper bound for the long-term stability).

The implication of this analysis is that the extrapolated worst-case Hg or Al ion clock stability (essentially its systematic uncertainty) will equal the stability of the extrapolated best millisecond pulsar stability at an integration time of the order of hundreds of years ($10^{10}$\,s): a measurement which would take many thousands of years of integration time to properly characterize.

The dimensionless ratio of two ideal unperturbed transition frequencies of different ions or atoms should be constant in the same reference frame (whether inertial or not) using the current standard model of particle physics.~\citep{Uzan} The ideal atomic clock attempts as closely as possible to yield an output frequency equal to that of an unperturbed transition and the accuracy evaluation is a combined theoretical and experimental test to quantify the difference between the actual output and the unperturbed value. It is of course always possible that there is some currently unknown physics beyond the standard model which will further perturb the transition and it is indeed this possibility which motivates current searches for failures in the standard model of physics using clock comparisons and atom interferometers. It is only meaningful to look for new physics in the variation of dimensionless ratios.~\citep{Uzan, Flambaum, Flambaum2}  To date however all tests aimed at searching for such new physics have yielded results consistent with the existing accuracy limits.  Nonetheless, one can accept that extreme long-term comparisons of timescales provided by ensembles of pulsars against that provided by atomic clocks might yield hints of some particular flaw in the current model of physics, which is not evident in comparisons of various atomic clocks, but which is seen because of the different physics which dominates the astrophysical sources.  
 
\begin{figure}[!t]
\centering
\includegraphics[width=3.5in]{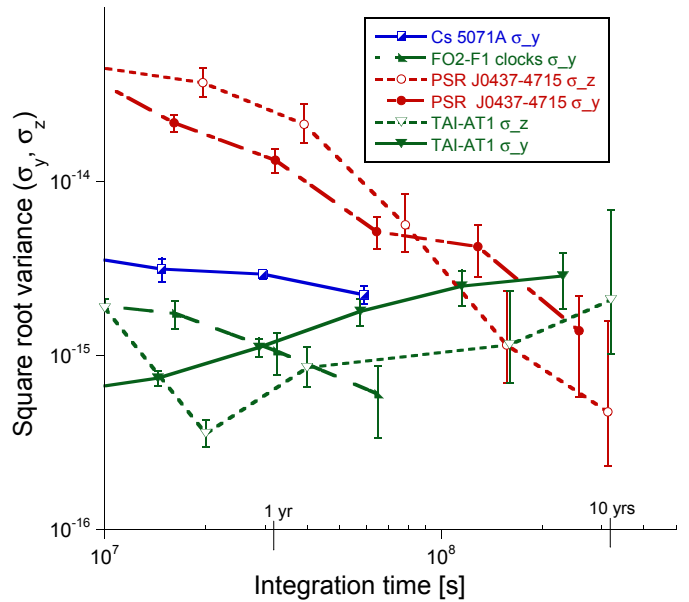}
\caption{\label{fig2} Derived frequency stability for some of the frequency sources of Fig. 1 for comparison over the longest timescales. Included are the pulsar PSR J0437-4715 (shown in red in online version),  a commercial cesium oscillator Cs 5071 (shown in blue) and some of the best laboratory clocks (shown in green).  The TAI-AT1 timescale comparison data and the pulsar/clock comparison data are represented with both the $\sigma_y$ and  $\sigma_z$  statistics.}
\end{figure}

\begin{table}
\begin{center}
\caption{Clocks and astrophysical sources and their period drift rates}
\begin{tabular}{lcccl}
\colrule\colrule
Source & $\dot{P}/P \, [s^{-1}]$ & Reference\\
\colrule
$^{199}Hg^+$ ion vs Cs clock 				& $1.3 \times 10^{-23}$ 	&~\citep{fortier2007}\\
$^{171}Yb^+$ vs Cs clock						& $1.4 \times 10^{-22}$ 	&~\citep{peik2004}\\
H $1\textit{S}-2\textit{S}$ vs Cs clock	& $2.0 \times 10^{-22}$ &~\citep{fischer2004}\\
\hline
PSR B1937+21 		& $6.74 \times 10^{-17}$ 	&~\citep{Rawley1987}\\ 
PSR J0437-4715 	& $9.95 \times 10^{-18}$ 	&~\citep{Hotan2006}\\
PSR J1909-3744 	& $4.76 \times 10^{-18}$ 	&~\citep{Hotan2006}\\
PSR J1713+0747 	& $1.86 \times 10^{-18}$  &~\citep{Hotan2006} \\
G117-B15A 			& $1.66 \times 10^{-17}$  &~\citep{Kepler2005} \\
ZZ Ceti, R548		&	$2.58 \times 10^{-17}$  &~\citep{Mukadam2003} \\
\colrule
\end{tabular}
\end{center}
\end{table}

\section{Period drift}
It appears in Kepler et al. (2005) that the authors have confused period drift rate $\dot{P}/P$ with the fractional variance of the period (which is numerically equal to the fractional variance of the frequency as defined in Eq.~(\ref{eq3}) above, i.e. $\dot{P}(t)/P(t) = -\dot{f}(t)/f(t)$). When using the corrected analysis presented here, it is interesting to revisit  the comparison of the drift rates of the pulse-to-pulse period of astrophysical sources 
against the drift rates of the best artificial clocks. 

In the circumstances where random walk of frequency is absent it is possible to obtain a stable estimate of the drift by fitting a line to a time sequence of frequency measurements.~\citep{barnes1971}  We have obtained these data for the best modern atomic clocks from measurements made in  connection with tests of the temporal stability of the universal constants~\citep{fortier2007,guena2008,rosenband2008} and summarized these in Table I.  It should be noted that these drift rates are obtained from the 1$\sigma$ uncertainty of  drift rates that are consistent with zero. Hence it can be seen that the upper limit of these drift rates for the best artificial clocks are five orders of magnitude below the lowest values for the pulsars.

The experimental measurement of the stability of white dwarf G117-B15A~\citep{Kepler2005}  should have suggested that the performance of the star as a clock was far from the state-of-the-art.  The authors indicate that they measured the epoch of the maximum pulse brightness to a resolution of the order of a second and then compared with the time delivered by a local time dissemination system (atomic timescale). As can be seen from Eq.~(\ref{eq3}), no normalization to period should be made in calculating the SRAV from timing errors, and keeping time to an accuracy of a second (even over 30 years) is far from the state-of-the-art in modern time dissemination systems.  Since the performance of the white dwarf pulses is inferior to that of the time dissemination system, it implies also that the stability and drift are also worse than the local time reference.

\section{$\sigma_y$ and $\sigma_z$ over the longest measurement times}
Power law fits to the calculated SRAV for the three pulsars in Fig.~1 have slopes very close to $1/\tau$, indicating white phase noise over these time scales.  The noise is dominated by signal to noise considerations in the receiver and time delays in the electronics, instrumental polarization, propagation delays in the interstellar plasma, and pulse phase jitter intrinsic to the pulsar.  It is possible that this can be improved somewhat by averaging longer~\citep{Hotan2006} and significant improvements in TOA precision can be made 
by paying close attention to the above mentioned effects, including improvements to 
receiver systems and construction of telescopes with larger collecting 
areas. 

However, it has been recently reported~\citep{Oates, Akatsuka} that in the relatively near future optical lattice neutral atom clocks and trapped ion clocks could potentially demonstrate a frequency stability of the order of $10^{-17}/\sqrt{\tau}$ were they to be able to overcome a challenging technical issue associated with their interrogating oscillators.  This would  make them more than two orders of magnitude better than the best terrestrial clocks shown on Fig.~1.   

Given that the best clocks and best pulsar observatories are unlikely to be co-located it is perhaps better to  consider that the uncertainties exemplified by the  comparison of fountain clocks located on different continents (FO2-F1 in Fig.~1 and 2) are the most fair representation of the level of performance to be expected from a state of the art clock with a high quality time transfer system.  The long-term performance of this complete system is limited by the systematic uncertainty of the Cs fountains which is at such a level that the stability of the astrophysical sources could only exceed that of  the joint clock-time transfer system for periods beyond a few hundred years of integration time.  

These clocks (FO2 and F1) were neither operated continuously nor simultaneously. Their comparison is based on individual comparisons with TAI, each lasting 15 to 30 days, and scattered during the nearly 11 year existence of the NIST  F1 fountain.  With the known properties of the underlying reference TAI one can indeed build a long term comparison. And, since time transfer noise averages much faster than white frequency noise, the time transfer noise is well below the limit of current state-of-the-art terrestrial clocks.  However the uncertainties associated with intercontinental comparisons still present a challenge and most researchers believe that recent work on time and frequency dissemination using optical fiber networks~\citep{Daussy2005, Kefelian2009} and comparisons to space-borne clocks~\citep{Uhrich2000} will result in orders of magnitude improvement to long-baseline clock comparisons in the very near future.

Finally a closer look is needed over the very longest times. For this reason we have expanded the scale of Fig.~1 around integration times of 1 to 10 years and only compare the best pulsar (which was measured against a terrestrial timescale) with terrestrial clock/clock and timescale/timescale comparisons that have data over that time period. This is shown in Fig.~2. 

Data from the TAI-AT1 timescale comparison and the millisecond pulsar PSR J0437-4715 are presented both as $\sigma_z$ and $\sigma_y$, the latter with period drift removed. The last $\sigma_y$ point for the PSR J0437-4715 data is determined from one half of the data set averaged with the other half: it is for this reason that the error bars are so large. The final $\sigma_z$ point  for the PSR J0437-4715 pulsar  and the TAI-AT1 timescale data comes from the length of whole times series data set. In the case of the pulsar that is 9.9 years~\citep{Verbiest2009} and in the case of the terrestrial timescales comparison the last 10 years were used in the analysis. The computed 1$\sigma$ error bars come from the limits placed on  $\sigma_z^2$ assuming it has a $\chi^2$ distribution with $n$ degrees of freedom,
where $n$ is the number of squared values of $c_3$ (the coefficient of the best fit cubic term in the time series data) appearing
in the average. However for the final few values the estimates of the $\sigma_z$ statistic will be biased low.~\citep{matsakis}

It should be noted that for the longest measured averaging times, $\sigma_y$ and $\sigma_z$ are essentially equivalent when the error bars are considered. It is clear that the quality of the data for this pulsar (which was measured against some terrestrial timescale) are insufficient to really separate the two measures. One sees that  for $\tau > 6$ years $\sigma_y$ of the pulsar and the TAI-AT1 comparison are also equivalent. This is has to be the case as the pulsar measurements are compared to a disseminated timescale as well, and so the results must be similar. The uncertainty in the timescale TT(BIPM2003), used by pulsar timing researchers, is $1 \times 10^{-15}$. \cite{TT(BIPM)}  The apparent slight differential between the  final $\sigma_z$ point for the pulsar and the final $\sigma_y$ values can be explained because of the increased rejection of the $\sigma_z$ measure of long term noise and the non-stationarity of the noise over these long term comparisons (see Fig. 5 in Ref.~\citep{matsakis96}). This is underlined by the final two $\sigma_z$ values for both the pulsar and the TAI-AT1 timescale comparison; within the error bars they are equivalent.

\section{Conclusion}
In conclusion, previous publications have indicated that some pulsars, or even white dwarfs, could, or do, provide the best clocks in the Universe. Although   measurements on time periods beyond seven years have yet to be performed on the most advanced terrestrial clocks, and there are insufficient data to properly characterize them beyond two years, the above analysis strongly implies that artificial clocks constructed within the time and frequency community are very likely to be better in terms of accuracy or stability over any accessible time period.  

Accuracy evaluations~\citep{rosenband2008, Lemke,  Chapelet, guena2008} on modern clocks imply that with longer measurements  their stability will get to at least the parts-in-$10^{17}$ regime. Natural clocks on the other hand would require measurement times over  many hundreds of years before they could confidently claim a frequency stability of this order. The trend line for $\sigma_y$ in Figs 1 and 2 or even for $\sigma_z$ in Fig.~2 indicates that it would require an ensemble (to eliminate short-term non-deterministic noise) of pulsars  averaged over many hundreds of years to reach  an \textit{rms}  stability level of parts in $10^{17}$, which is an upper limit on the most accurate terrestrial clocks. 

The proviso on this conclusion is that advanced terrestrial clocks have not yet been measured over periods greater than about 6 years, which means statistics for periods greater than 2 years are not yet available, nor have time dissemination systems operated that will allow this stability to be broadly available globally (something that is equally crucial for comparison of astrophysical sources using these advanced terrestrial clocks and timescales including them).  On the latter point we note active development in this field has already successfully demonstrated sub-$10^{-15}$ accuracy transfer over hundreds of kilometers.~\citep{Kefelian2009} On the former point we note that the  accuracy and stability of terrestrial clocks have improved more than an order of magnitude, on average, in each decade over the last 60 years, since the development of the first atomic Cs clock, while the timing stability of the best millisecond pulsar detected by comparison with local atomic clocks  has improved by less than an order of magnitude in the last 20 years.\cite{footnote3}  Recently identified paths~\citep{Oates, Akatsuka} to further improved terrestrial clocks suggests  that this trend will continue  into the future. 

\acknowledgments
We are grateful to  Matthew Bailes and Joris Verbiest for the electronic data for some millisecond pulsars (the timing residuals for the three pulsars PSR J0437-4715, PSR J1909-3744 and PSR J1713+0747 and $\sigma_z$ for PSR J0437-4715) and to Tom Parker for NIST AT1 free running ensemble data,  Cs 5071 data, and the comparison between SYRTE and NIST fountains FO2-F1 data. Also we would like to thank Dick Manchester and Tom Parker for valuable discussions.

\clearpage


\begin{thebibliography}{99}
\bibitem{footnote1} Technically these clocks cannot be more accurate than cesium primary frequency standards that currently define the second. In order to realize this accuracy it would require a redefinition of the second.
\bibitem{footnote2} We note proposals to monitor many pulsars simultaneously, which can eliminate this dependence on terrestrial timescales, or at least provide a flywheel to correct atomic time over very long time periods.
\end{thebibliography}
\end{document}